# Experimental observation of superscattering

Chao Qian[1,2,†], Xiao Lin[3,†,*], Yi Yang[4], Xiaoyan Xiong[2], Huaping Wang[1], Erping Li[2], Ido Kaminer[5,6], Baile Zhang[3,7,*] and Hongsheng Chen[1,2,*]

[1]*State Key Laboratory of Modern Optical Instrumentation, The Electromagnetics Academy at Zhejiang University, Zhejiang University, Hangzhou 310027, China.*
[2]*ZJU-UIUC Institute, College of Information Science and Electronic Engineering, Zhejiang University, Hangzhou 310027, China.*
[3]*Division of Physics and Applied Physics, School of Physical and Mathematical Sciences, Nanyang Technological University, Singapore 637371, Singapore.*
[4]*Department of Electrical Engineering and Computer Science, Massachusetts Institute of Technology, Cambridge, MA 02139, USA.*
[5]*Department of Electrical Engineering, Technion - Israel Institute of Technology, Haifa 32000, Israel*
[6]*Department of Physics, Massachusetts Institute of Technology, Cambridge, MA 02139, USA.*
[7]*Centre for Disruptive Photonic Technologies, Nanyang Technological University, Singapore 637371, Singapore.*
[†]*These authors contributed equally to this work.*
[*]*Corresponding authors. Email:* xiaolinbnwj@ntu.edu.sg *(X. Lin);* blzhang@ntu.edu.sg *(B. Zhang);* hansomchen@zju.edu.cn *(H. Chen)*

**Abstract:** Superscattering, induced by degenerate resonances, breaks the fundamental single-channel limit of scattering cross section of subwavelength structures; in principle, an arbitrarily large total cross section can be achieved via superscattering. It thus provides a unique way to strengthen the light-matter interaction at the subwavelength scale, and has many potential applications in sensing, energy harvesting, bio-imaging (such as magnetic resonance imaging), communication and optoelectronics. However, the experimental demonstration of superscattering remains an open challenge due to its vulnerability to structural imperfections and intrinsic material losses. Here we report the first experimental evidence for superscattering, by demonstrating the superscattering simultaneously in two different frequency regimes through both the far-field and near-field measurements. The underlying mechanism for the observed superscattering is the degenerate resonances of confined surface waves, by utilizing a subwavelength metasurface-based multilayer structure. Our work paves the way towards practical applications based on superscattering.

Enhancing the electromagnetic scattering from subwavelength nanostructures [1-4] can enable a physically small object to capture wave energy from a large area, and is thus of great importance to many practical applications [5-10], such as for antenna design, energy harvesting, biomedical imaging (e.g., magnetic resonance imaging, MRI), and communications. To characterize the scattering response, typically either the absolute [11,12] or the area/volume-normalized [13-15] scattering cross sections are chosen as the figure of merit, depending on the application scenarios. In terms of the absolute scattering cross section, there is a constraint on achieving strong scattering with subwavelength scatterers. For a subwavelength scatterer (e.g., an atom) in a two-dimensional (2D) homogeneous environment with a refractive index of $n$, one can rigorously prove that its maximum absolute cross section cannot exceed $2\lambda/n\pi$ [16], where $\lambda$ is the wavelength of light in free space. This limit is referred as the single-channel limit, and a similar limit ($3\lambda^2/2n\pi$) exists for the 3D case [17]. The phenomenon of superscattering, induced by degenerate polaritonic resonance, is revealed to overcome the aforementioned single-channel limit by the pioneering work in 2010 [11]. In principle, the superscattering can enable an arbitrary large absolute total cross section from subwavelength structures, if one maximizes contributions from a sufficient number of channels [11,12]. Therefore, given the enticing application prospects, the superscattering is highly desirable if experimentally realized.

However, it remains an open challenge to experimentally observe superscattering. The challenge mainly stems from the complexity of structural fabrication and the detrimental influence of realistic material losses. For example, to directly observe superscattering, a subwavelength multilayer structure is necessary to create degenerate resonances of confined transverse magnetic (i.e., TM, or *p*-polarized) polaritons, such as surface plasmon polaritons in metal [11,12] or graphene [18] and phonon polaritons in hexagonal boron nitride (BN) [19-21]. This involves the fabrication of *free-standing* multilayer micro/nano-structures, with the precise requirement on the thickness of each layer, and thus increases fabrication complexity. Moreover, when realistic material losses are taken into account, the superscattering phenomenon will deteriorate and even disappear [22]. Due to the above challenges, the superscattering was *only* widely theoretically studied, but remains experimentally elusive.

In this work, we for the first time experimentally observe the phenomenon of superscattering from a subwavelength structure, through the far-field measurement of scattering cross section and the near-field measurement of scattered fields in the microwave regime. The subwavelength structure, constructed by three conformal and dielectric-separated metasurfaces rolled in a cylinder shape [see Fig. 1(a)], supports degenerate resonances of confined transverse electric (TE, or *s*-polarized) surface waves [23-26]. Importantly, we observe degenerate resonances simultaneously in two different frequency ranges, verifying the prediction of *multifrequency* superscattering [19]. These pronounced observations on superscattering mainly rely on the flexibility in fabrication of macroscopic structures in the microwave regime and the negligible propagating loss of confined TE surface waves in the designed metasurfaces [27-30]. Our work also shows that the superscattering can be generalized to TE waves in nonmagnetic systems, going beyond previous theoretical works of superscattering that focused on TM waves [11,12,18,19]. As a result, superscattering can be achieved without stringent requirements on the polarization of light and can benefit applications relying on TE waves [5-10].

*Multifrequency superscattering of TE waves from a subwavelength structure.*—For conceptual clarity, we consider the superscattering of TE waves in 2D from a subwavelength circular scatterer. The circular scatter in 2D is equivalent to an ideal 3D rod of infinite length along z-direction, and can be effectively realized by a 3D rod of finite length $L$, which is much larger than the studied wavelength $\lambda$ [19], i.e., $L \gg \lambda$; see Fig. 1(a). The subwavelength rod is composed of three conformal and dielectric-separated metasurfaces. The ultrathin metasurfaces (with thickness $d \ll \lambda$) are made of deep-subwavelength periodic copper strips, which are separated by air grooves of width $g$. If the pitch $w$ of copper strips is also much smaller than the interested wavelength, i.e., $w \ll \lambda$, the metasurface can be reasonably described by a surface conductivity $\sigma_s$ [31,32]. The value of $\sigma_s$ is pure imaginary and negative in the microwave regime, i.e., $\sigma_s = iIm(\sigma_s)$ and $Im(\sigma_s) < 0$; see Supplemental Note 1. From the pioneering work of TE surface waves in *nonmagnetic* systems in 2007 [23], each metasurface with a surface conductivity of $Im(\sigma_s) < 0$ supports TE surface waves, while TM surface waves appear only when $Im(\sigma_s) > 0$. Moreover, the TE surface waves can be highly confined in space if the value of $|Im(\sigma_s)|$ is sufficiently large [25,26], i.e., $|Im(\sigma_s)| \gg G_0$, where $G_0 = e^2/4\hbar$ is the universal optical conductivity. For the designed metasurfaces, we have $|Im(\sigma_s)| > 80G_0$ in the studied frequency range; see Fig. S3. Therefore, the designed subwavelength metasurface-based rod supports multiple eigenmodes of *confined* TE surface waves within the interested frequency range; see Fig. S4. This provides the possibility to realize superscattering with confined TE surface waves at one or even multiple frequencies through structural optimization, as shown in Fig. 1.

As depicted in Fig. 1(a), the optimized subwavelength rod has a diameter of $D = 35.96$ mm. In the cross-sectional view of the structure, from inside towards outside, the dielectric region has a thickness of $d_1 = 3.4$ mm, $d_2 = 1.7$ mm, $d_3 = 6.2$ mm and $d_4 = 6.68$ mm, with a relative permittivity of $\varepsilon_{r1} = 2.4$, $\varepsilon_{r2} = 2.7$, $\varepsilon_{r3} = 2.3$ and $\varepsilon_{r4} = 2.1$, respectively. These dielectrics are implemented with the polytetrafluoroethylene material with the required value of permittivity. In our experiments, the designed rod is placed in free space, which is denoted as region 5 with $\varepsilon_{r5} = 1$. The surface conductivities of the designed three different metasurfaces are $\sigma_{2|3}$ (i.e., the metasurface at the interface between regions 2&3), $\sigma_{3|4}$ and $\sigma_{4|5}$, respectively; see Fig. 1(a). There is no metasurface at the interface between regions 1&2 in our design; we simply denote $\sigma_{1|2} = 0$. For these metasurfaces, from inside towards outside, the pitch of periodic copper strips are $w_{2|3} = 11.29$ mm, $w_{3|4} = 16.94$ mm and $w_{4|5} = 33.88$ mm and the widths of air grooves are $g_{2|3} = g_{3|4} = 0.14$ mm and $g_{4|5} = 0.1$ mm, respectively. These metasurfaces are closely attached to the dielectric interfaces; see Fig. 1(a). The length of the designed rod is $L = 510$ mm, fulfilling the condition of $L \gg \lambda$.

The scattering cross section of TE waves in 2D from the multilayer rod can be solved analytically, based on the Mie scattering theory [33]. In short, the total scattering cross section, i.e., the total scattered power over the intensity of incident waves, can be expressed by $C_{sct} = \sum_{m=-\infty}^{\infty} C_{sct,m}$, where $C_{sct,m} = \frac{2\lambda}{\pi}|S_m|^2$. The parameter $S_m$ is the scattering coefficient of the $m^{\text{th}}$ angular momentum channel and can be calculated by matching the boundary conditions; see Supplemental Note 1. Since $|S_m| \leq 1$, the scattering cross section $C_{sct,m}$ from individual channel is bounded from above by the single-channel limit $\frac{2\lambda}{\pi}$, i.e., $C_{sct,m} \leq \frac{2\lambda}{\pi}$ [16,17]. Due to the consistent degeneracy of the $m^{\text{th}}$ and $-m^{\text{th}}$ angular momentum channels (i.e., $C_{sct,m} = C_{sct,-m}$), $C_{sct,|m|}$ is less than two times of the value of single-channel limit, i.e., $C_{sct,|m|} \leq \frac{4\lambda}{\pi}$. Superscattering goes beyond such a limit by creating decoupled degenerate resonances [11] from channels with *different* nonzero angular momenta $|m|$ (such as $|m| = 1$ and 2 in Fig. 1(b)). In order to highlight the scattering cross section from the polaritonic resonance (i.e., $|m| \geq 1$), below we mainly consider the scattering cross section from the channels with $m \neq 0$.

Figure 1(b) shows the analytical scattering cross section of TE waves from the designed subwavelength multilayer rod, after structural optimization by simulated annealing algorithm [34]. The total scattering cross section is at least 4 times of the value of single-channel limit at both 2.2 GHz ($\lambda = 135.7$ mm) and 3.73 GHz ($\lambda = 80.4$ mm). Therefore, the diameter 35.96 mm of the designed multilayer rod is 0.26 and 0.45 times of the wavelengths at the above two frequencies, respectively. The scattering cross sections from the $|m| = 1$ or 2 channels reach the single-channel limit at these two resonant frequencies; see Fig. 1(b). These maximum single-channel scattering cross sections all originate from the resonances of the confined TE surface waves, as evident from the field distribution of scattered waves from different channels in Fig. S2. Moreover, the degenerate resonances of the confined TE surface waves appear simultaneously at these two frequencies, which lead to the multifrequency superscattering [16]. In the following, the designed subwavelength metasurface-based rod is thus denoted as the superscatterer.

To experimentally obtain the scattered waves (in the far or near field), we implement measurements in two sequential steps. First, we measure the incident field without the scatterer. Second, we measure the total field with the scatterer. Due to the advance of microwave technology, we can simultaneously obtain the amplitude and phase for these measured fields [35]. The scattered field can be obtained by subtracting the incident field from the total field; the scattering cross section can be calculated accordingly.

*Far-field measurement of scattering cross section.—*Figure 2(a) shows the measured total scattering cross section of TE waves from the superscatterer, confirming the occurrence of multifrequency superscattering; see the far-field measurement setup in Fig. S5a. The measured total scattering cross sections at 2.2 GHz and 3.73 GHz exceeds four times the single-channel limit. The measured results are consistent with the analytical results for the ideal circular structure in 2D as shown in Fig. 1(b) and the simulated results for the real 3D structure through the software of CST Microwave Studio. Such excellent agreement mainly comes from the validity of the perfect electric conductor (PEC) treatment for the copper strips in the microwave regime, as well as the negligible propagation loss of the confined TE surface waves in the designed copper-based metasurfaces. The minor deviations between the measured and analytical/simulated results are mainly due to the structural imperfection during fabrication and the small deviation from the designed permittivities from the chosen dielectrics. For comparison, Fig. 2(b) shows the total scattering cross section from a homogeneous PEC rod, with a same diameter (i.e., same geometrical cross section) as the superscatterer. For the PEC rod, the total scattering cross section is below two times of the value of single-channel limit (i.e., $C_{sct,m=1} = C_{sct,m=-1} \leq \frac{2\lambda}{\pi}$), indicating the absence of superscattering---degenerate resonances across different angular momenta $|m|$.

Figure 3 shows the measured far-field radiation pattern of TE waves from the superscatterer. Here we use the cross section per azimuthal angle $\frac{\partial C_{sct}}{\partial \phi} = 2\pi\rho_0 \frac{|E_s|^2}{|E_i|^2}$ (i.e., the scattering width [36]) as a function of the azimuthal angle $\phi$ to characterize the far-field radiation pattern for the structure in 2D. Here $\phi$ is the angle between the wavevector $\bar{k}$ of scattered waves and the $\hat{x}$ coordinate, since we assume the incident TE plane waves propagate along the $+\hat{x}$ direction. $E_i$ and $E_s$ are the incident and scattered fields measured at $\rho = \rho_0$, respectively, and $\rho$ is the radial distance between the measured scatterer (i.e., the superscatterer or PEC rod) and the detector (an antenna). In our experiments, $\rho_0 = 1.5$ m (having $\rho_0 > 10\lambda$) is adopted to guarantee that the measurement is carried out in the far-field [37]. After obtaining the cross section per azimuthal angle, the total scattering cross section can be calculated via $C_{sct} = \frac{1}{2\pi}\int_0^{2\pi} \frac{\partial C_{sct}}{\partial \phi} d\phi$. Taken together, Figs. 3(a,b) show that the superscatterer has the maximum cross section per azimuthal angle in the forward direction, i.e., at $\phi$ close to 0°, for both superscattering frequencies. Moreover, in the forward direction, the cross section per azimuthal angle from the superscatterer is much larger than that from the PEC rod in Figs. 3(c) and 3(d). In addition, from the measured scattered fields, we can further calculate the scattering coefficients $S_m$ for different $m$, which are $|S_{\pm 1}| = 0.89$ and $|S_{\pm 2}| = 0.94$ at 2.2 GHz, $|S_{\pm 1}| = 0.92$ and $|S_{\pm 2}| = 0.97$ at 3.73 GHz, respectively; see details in Supplemental Note 4. These measured scattering coefficients are close to unity -- consistent with the theoretical values. This further verifies that the observed multifrequency superscattering in Fig. 2 is due to the appearance of decoupled degenerate resonance.

*Near-field measurement of the scattered fields.*—Figure 4 shows the measured spatial distribution of scattered waves in the near field; see the experimental setup in Fig. S5(b). The TE plane waves are incident from the left side. Notably, the near-field distribution of scattered waves at both superscattering frequencies demonstrate the mixed pattern between the dipole and quadrupole resonances; moreover, the scattered waves from the superscatterer mainly appear in the forward direction [see Figs. 4(a, c, e, g)], which are much stronger than those from the PEC rod [see Figs. 4(b, d, f, h)]. These phenomena are because the scattering fields from channel $|m| = 1$ and from channel $|m| = 2$ are in-phase (out-of-phase) in the forward (backward) direction [19], leading to the constructive interference of scattering fields from different channels in the forward direction. Consequently, the incident plane waves are largely disturbed at both superscattering frequencies, and a remarkable "shadow" leaves behind the superscatterer; see Fig. S1. The results in Fig. 4 offer complementary evidence for the degenerate resonance-induced superscattering, aside from the large cross section shown in Figs. 2 and 3.

*Discussion.—*In conclusion, we have directly observed the superscattering in 2D from a subwavelength circular rod, confirmed by both the far-field and near-field measurements at the microwave frequency regime. Our work thus facilitates the potential observation of superscattering in 3D (e.g., from a subwavelength sphere) and even in other high frequency ranges, and many new concepts for applications based on superscattering, that can be applied for single molecule fluorescence imaging [8], optical tagging [5,9] and energy harvesting [10]. In addition to the superscattering, there are other schemes to enhance the scattering from tiny objects, including the use of enhanced directivity [38], negative index materials for the design of scatterers [39], homogeneous but low index environment [40] and even inhomogeneous environment [41]. It is noted that there are rare studies that combine the scheme of superscattering with other schemes to further enhance the scattering from tiny objects. For example, since the single channel limit $2\lambda/n\pi$ is inversely proportional to $n$, the scattering response in the environment with $n \to 0$ shall be much stronger than that in free space with $n = 1$. However, the phenomenon of superscattering in the low index environment has not been investigated yet, even in theory. The continuing pursuit of such exotic scattering phenomena is still highly desired, since they may further extend our capability to flexibly manipulate the light-matter interaction in the extreme nanometer scale.

**Acknowledgements**
This work was sponsored by the National Natural Science Foundation of China under Grant Nos. 61625502, 61574127, 61601408, 61775193, and 11704332, the ZJNSF under Grant No. LY17F010008, the Top-Notch Young Talents Program of China, the Fundamental Research Funds for the Central Universities, the Innovation Joint Research Center for Cyber-Physical-Society System, the Singapore Ministry of Education (Grant No. MOE2015-T2-1-070, MOE2016-T3-1-006, and Tier 1 RG174/16 (S)). I. Kaminer is an Azrieli Fellow, supported by the Azrieli Foundation, and was partially supported by the Seventh Framework Programme of the European Research Council (FP7-Marie Curie IOF) under grant no. 328853-MC-BSiCS.

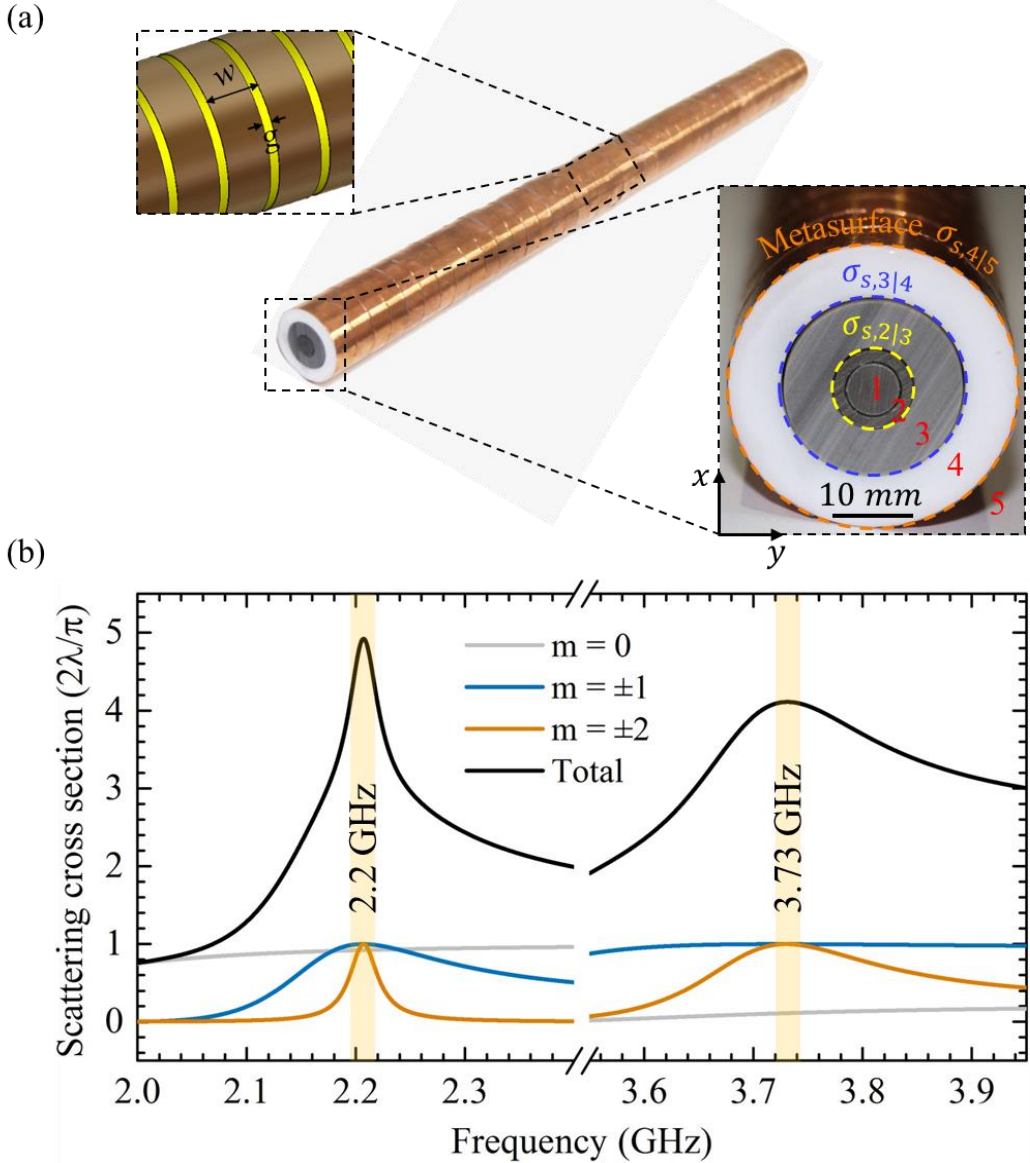

FIG. 1. Multifrequency superscattering of TE waves from a subwavelength structure. (a) Photograph of the fabricated subwavelength multilayer circular rod. Here the realistic 3D rod with its finite length much larger than the interested wavelength can be reasonably treated as an ideal rod in 2D (with an infinite length). The rod is composed by three conformal metasurfaces which are parallel to each other; see colored dashed lines in the right inset. The ultrathin metasurface, constructed by deep-subwavelength periodic copper strips (see the left inset), can be theoretically described by a lossless surface conductivity of $\sigma_s$ with $\text{Im}(\sigma_s) < 0$. The designed metasurfaces support TE surface waves, instead of TM surface waves which require $\text{Im}(\sigma_s) > 0$. (b) Analytical total scattering cross section and the contribution from individual channel, from the designed circular rod in 2D located in free space. The degenerate resonance of confined surface waves emerges simultaneously at 2.2 GHz and 3.73 GHz, leading to the multifrequency superscattering.

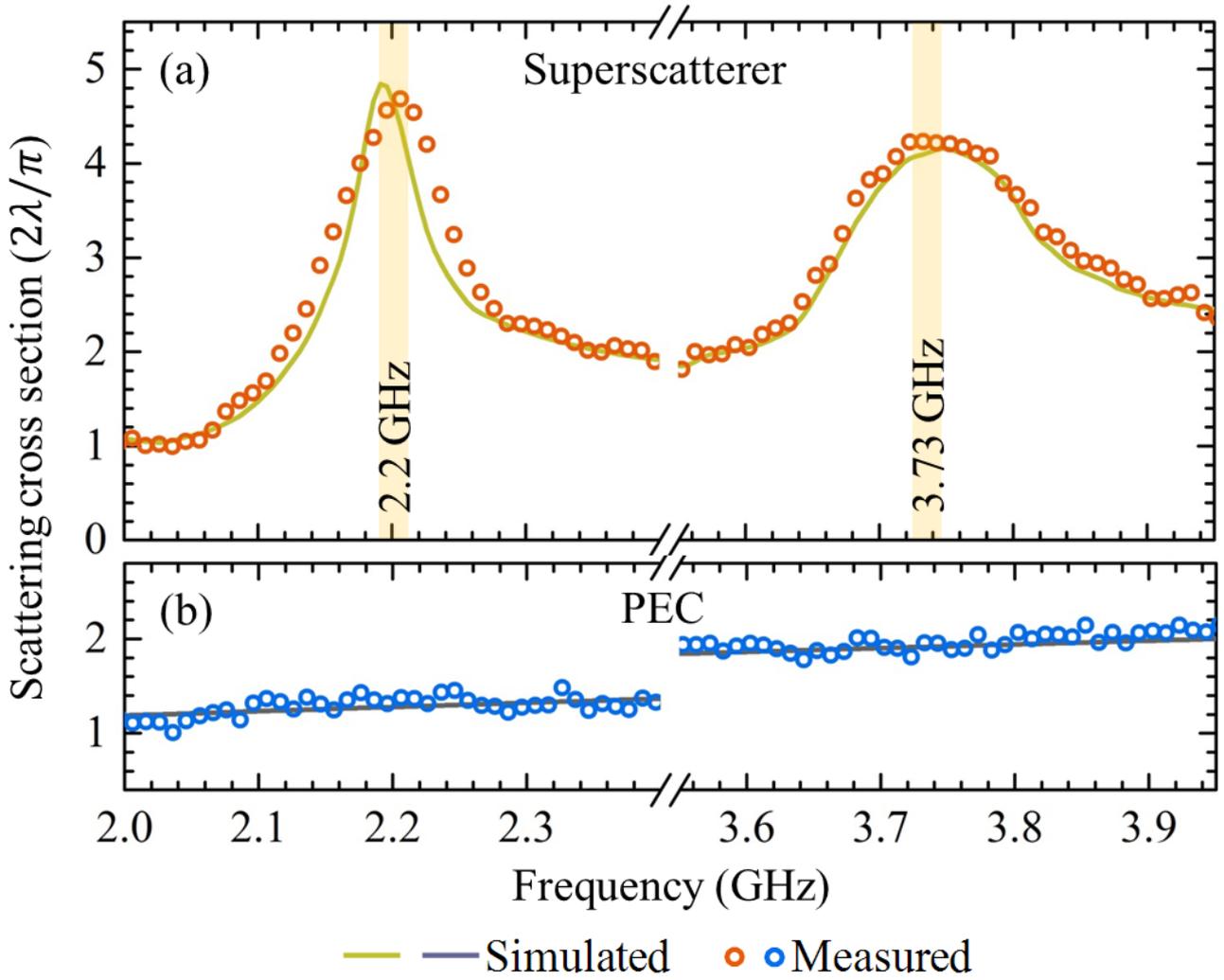

FIG. 2. Far-field measurement of total scattering cross section of TE waves (a) from a subwavelength multilayer structure (denoted as superscatterer) and (b) from a circular perfect electric conductor (PEC) rod. The superscattering of the designed superscatterer appears simultaneously at 2.2 GHz ad 3.73 GHz. The PEC rod has a same diameter with the superscatterer which is shown in Fig. 1(a). The measured results (colored dots) are consistent with the simulated results for the real 3D structure by software CST (colored lines) and the analytical results for the ideal structure in 2D as shown in Fig. 1(b).

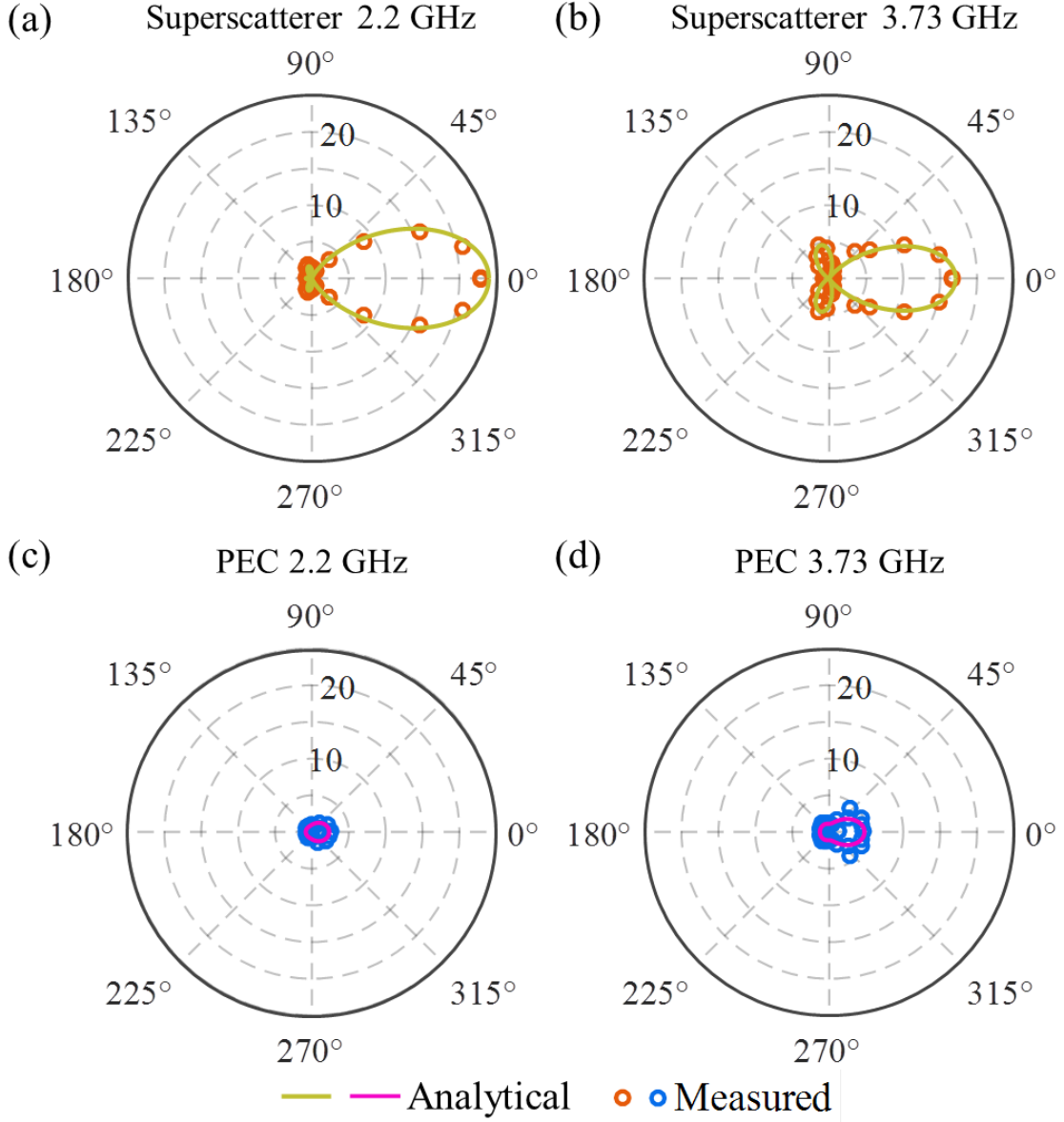

FIG. 3. Far-field measurement of the scattering cross section per azimuthal angle of TE waves (a, b) from the superscatterer and (c, d) from the PEC rod. 2.2 GHz and 3.73 GHz are the two superscattering frequencies for the designed superscatterer. The cross section per azimuthal angle is shown as a function of the azimuthal angle $\phi$, where $\phi$ denotes the angle between the $\hat{x}$ coordinate and the wavevector $\bar{k}$ of scattered waves. The incident waves propagate along the $+\hat{x}$ direction, i.e. $\phi = 0°$. The radial length represents the magnitude of cross section per azimuthal angle, which is in unit of $2\lambda/\pi$. The measured results (colored dots) are consistent with the analytical results (colored lines).

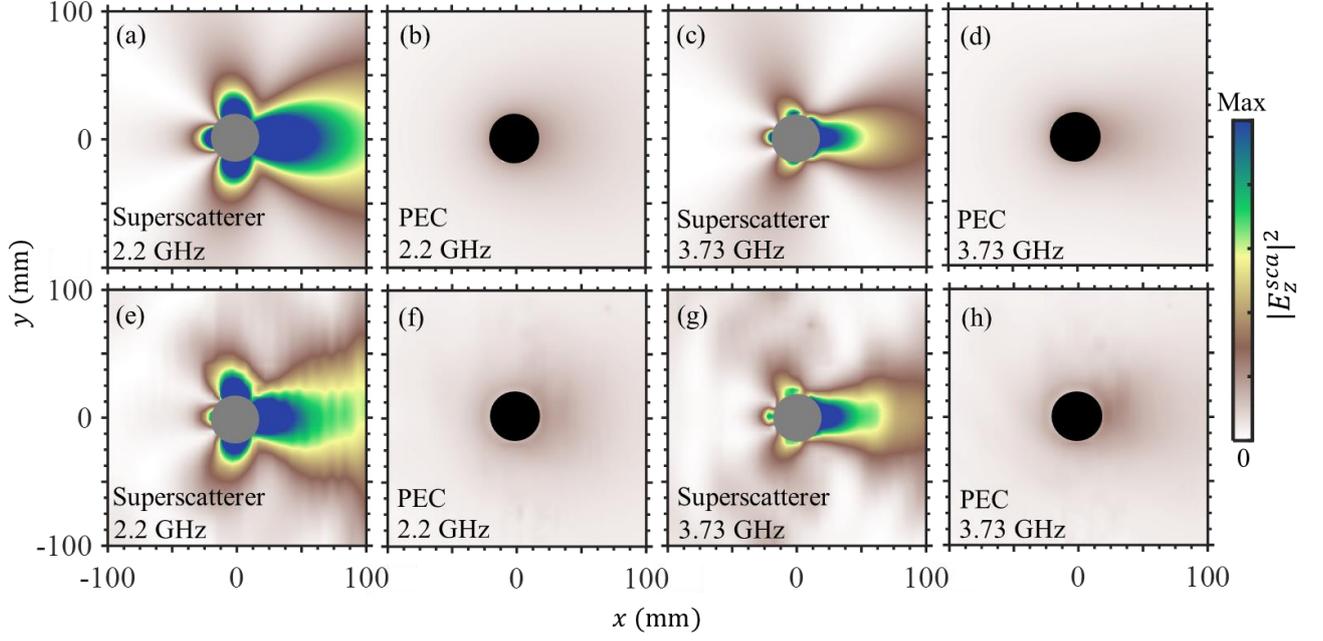

FIG. 4. Near-field measurement of the scattered field $|E_z^{sca}|^2$ (a, c, e, g) from the superscatterer and (b, d, f, h) from the PEC rod. Much pronounced scattered fields are observed in the superscatterer. The measured results in (e-h) are consistent with the analytical results in (a-d). The designed superscatterer has the superscattering phenomenon simultaneously at 2.2 GHz and 3.73 GHz. The incident plane waves come from the left side and propagate along the $+\hat{x}$ direction. The superscatterer and the PEC rod are highlighted by gray and black circles, respectively.

13